\documentclass{appolb}
\usepackage{epsfig}

\begin{document}
\title{SEARCH FOR $\eta$-MESIC HELIUM\\USING THE WASA-AT-COSY DETECTOR%
\thanks{Presented at the Symposium on Meson Physics, Cracow, 01-04 October 2008.}%
}
\author{
Wojciech~Krzemie\'n$^{\star,\$}$$^,$\footnote{E-mail address: wojciech.krzemien@if.uj.edu.pl}~,
Pawe{\l}~Moskal$^{\star,\$}$, Jerzy~Smyrski$^{\star}$\\ on behalf of the WASA-at-COSY collaboration
\address{
$^{\star}$ Institute of Physics, Jagiellonian University, Cracow, Poland\\
$^{\$}$ Institut f\"ur Kernphysik and J{\"u}lich Center for Hadron Physics, \\Forschungszentrum J\"ulich, J\"ulich, Germany \\
}
}
\maketitle
\begin{abstract}
We conduct a search for the $^4{\mbox{He}}-\eta$ bound state with the WASA-at-COSY facility via the measurement
of the excitation function for the reaction $dd \rightarrow {^3\mbox{He}}\, p\, \pi^-$. 
In first experiment performed in June 2008, we used COSY deuteron 
beam with a slowly ramped beam momentum corresponding to a variation of the excess energy 
for the $^4$He$\eta$ system from -51.4~MeV to 22~MeV.
Here we report on the status of the measurement and the data evaluation.
\end{abstract}
  
\section{Introduction}
Based on the fact that the interaction between the $\eta$ meson and nucleon is attractive, Haider and Liu postulated the existence of a new kind of matter in form of $\eta$-nucleus bound states~\cite{liu2}.
Because the $\eta$ meson is neutral such a system can be formed only via the strong interaction which distinguishes it qualitatively from exotic atoms where the
meson is bound via the electromagnetic force. 
Therefore the search for a signature of an $\eta$-mesic nucleus is interesting on its own account, 
additionally the experimental determination of the width and binding energy 
of $\eta$-mesic nuclei would provide a valuable input for studies of the $\eta-N$ interaction. 
It would provide information about properties of the $N^*(1535)$
embedded in
nuclei, the behavior of the $\eta$ meson
in the nuclear medium~\cite{osetNP710}, and for the determination of the flavor singlet component 
of the $\eta$ meson~\cite{bass}.
The formation of a bound state can only take place in nuclei, for
which the real part of the $\eta$--nucleus scattering length is negative
(attractive nature of the $\eta$--nucleus interaction), and the
modulus of the real part of $\eta$--nucleus scattering length is greater than the
modulus of its imaginary part~\cite{liu3}:
\begin{equation}
|Re(a_{\eta-nucleus})|>|Im(a_{\eta-nucleus})|.
\label{11}
\end{equation}
The relatively small value of the s-wave $\eta$N
scattering length known in 1980's ($a_{\eta N}~=~(0.28+0.19i)$~fm~\cite{liu1}) 
implied the possibility of forming an $\eta$--mesic nuclei only for A~$\ge$~12~\cite{liu2}. 
This estimation was strengthened by calculations of Li~\cite{li}.
However, recent theoretical considerations of hadronic- and photoproduction
of the $\eta$ meson result in a wide range of possible values of the
$\eta$--nucleon s-wave scattering lengths from $a_{\eta N}~=~(0.27+0.22i)$~fm up
to $a_{\eta N}~=~(1.05+0.27i)$~fm, with the suggested average value of $a_{\eta N}~=~(0.5+0.3i)$~fm.
Such a high value of the $\eta$--nucleon scattering length may enable the formation
of a bound $\eta$--nucleus states in light nuclei region such as $^{3,4}{\mbox{He}}$~\cite{wilkin1,wycech1}
and even in the deuteron~\cite{green}.
According to the calculations including multiple scattering theory~\cite{wycech1}
and Skyrme model~\cite{scoccola} an especially good candidate for binding is the $^4{\mbox{He}} -\eta$ system. 
It should be stated that at present there is no univocal picture of this issue
and results presented in article~\cite{osetPLB550} indicate that the ratio 
of the predicted width to the binding energy  becomes bigger for lighter nuclei 
which can make an observation of the bound state in light nuclei more difficult.
However, there are promissing indirect experimental observations
which may be interpreted as an indications of $\eta$-helium bound states.
For example analysis of the data from the close to threshold measurements
 of the total cross section for the $dp\to ^{3}\!{\mbox{He}}\eta$ reaction
by SPES-4~\cite{berger} and SPES-2~\cite{mayer} collaborations suggested 
a possible existence of the  bound
$^{3}{\mbox{He}}-\eta$ system since within the experimental errors the condition given in Eq.~\ref{11} could be  fulfilled.
A search for a $\eta$--nucleus bound state has also been performed in the hadronic channel at the
cooler synchrotron COSY, where the COSY-11 and ANKE collaborations independently,
using different detection setups, performed measurements of the
excitation function and differential cross sections for the $dp\to ^{3}\!{\mbox{He}}\eta$ reaction in the vicinity
of the kinematical threshold~\cite{jurek-he3,adam,timo}.
Both groups used the momentum ramping technique of the beam
deuterons in order to reduce the systematic errors. Measurements have been performed with beam momenta varying from
below the reaction threshold, up to an excess energy of about 8.5~MeV in the case of
the COSY-11 experiment and about 11.5~MeV in the case of the ANKE experiment.
Data taken by the COSY-11 group  were used to search for a signal of a
$^3{\mbox{He}}-\eta$ bound state below the $\eta$ production threshold, via the $dp\to ppp\pi¯$ and $dp\to ^{3}\!{\mbox{He}}\pi^0$ reactions~\cite{jurekaps,jurekmeson08,jurek2},
while the measurements above the threshold enabled the study of the forward-backward
asymmetries of the differential cross sections and the extraction of the $\eta ^{3}{\mbox{He}}$ scattering length.
The data of both groups~\cite{jurek-he3,timo} shows a variation in the 
phase of the s-wave amplitude in the near-threshold
region, consistent
with possible existence of a bound state~\cite{wilkin2,wilkinaip}.

The first direct experimental indications of a light $\eta$-nucleus bound system was
reported by the TAPS collaboration in the $\eta$  photoproduction 
reaction
$\gamma {^3\mbox{He}} \rightarrow \pi^0 p X$ ~\cite{mami}.  There the
difference between excitation functions for two ranges 
of the $\pi^0 - p$ relative
angle in the center-of-mass frame revealed a structure which was interpreted
as a possible signature of a bound state.
However, it was claimed that due to the limited statistics the result
could also be interpreted as an evidence for a virtual state~\cite{Han}.
Therefore a new, statistically more significant experimental confirmation is needed.

\section{Experimental method and decay model}

\begin{figure}[h]
\psfig{file=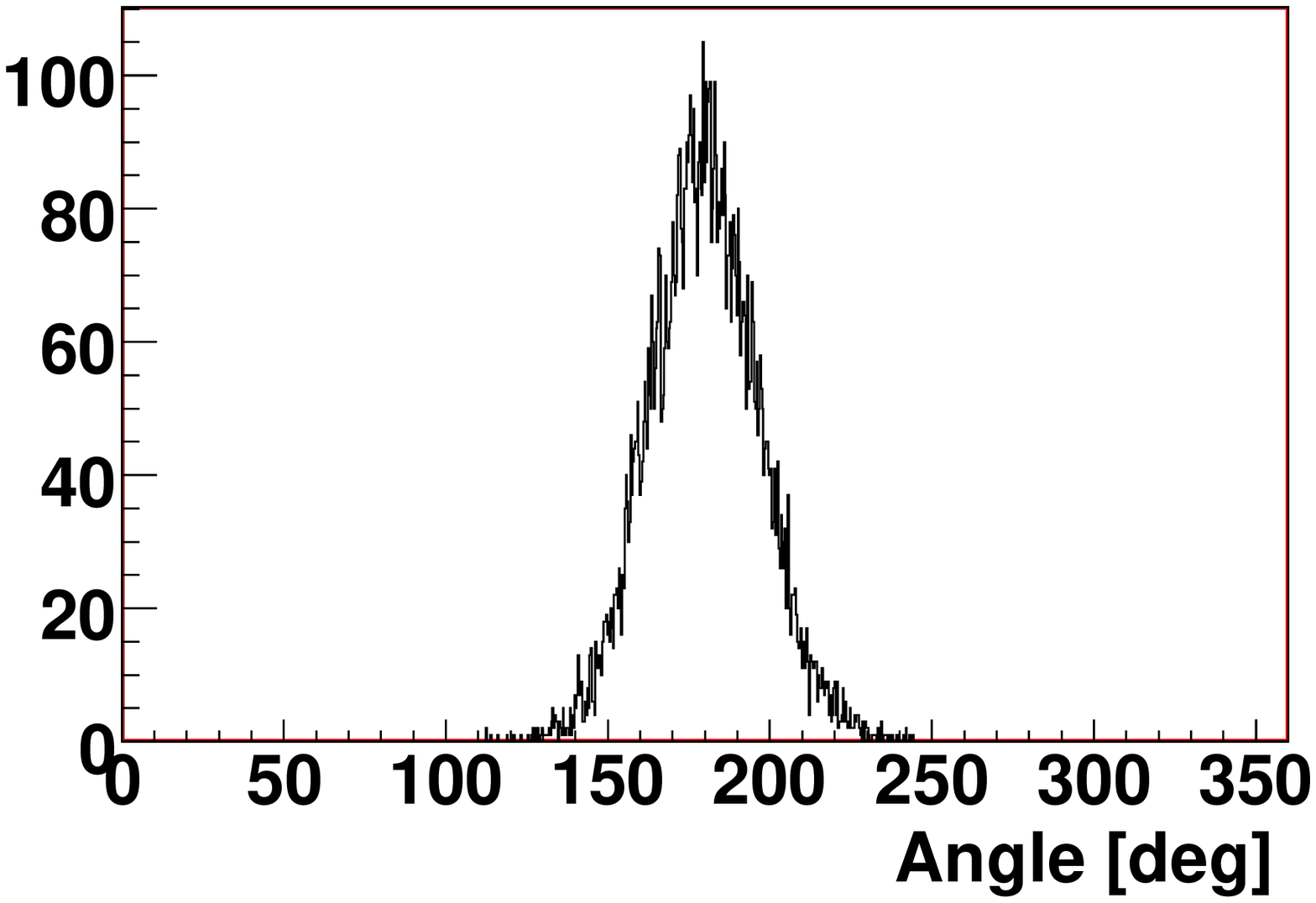,width=6.1cm}
\psfig{file=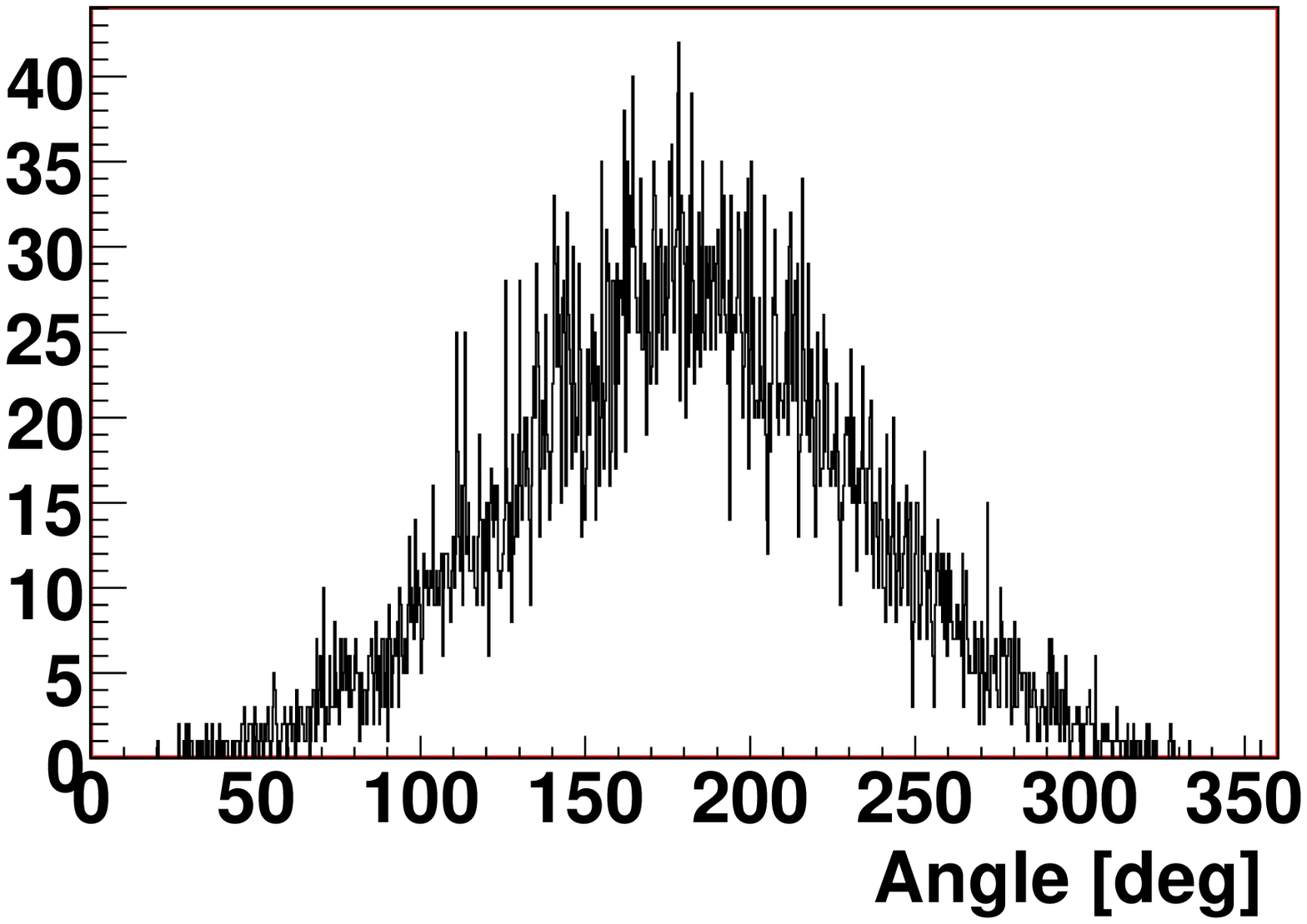,width=6.1cm}
\caption{\label{angle} 
Distribution of the relative $p-\pi$  angle seen in the reaction center-of-mass system
as simulated for the processes leading to the creation of $\eta$-helium bound state:  $dd\to(^{4}{\mbox{He}}\eta)_{bound}\to^3{\mbox{He}}p\pi$ (left),
and for the prompt production of the $^3{\mbox{He}}p\pi$ system assuming a homogeneous
population of the phase space for the  $dd\to^3{\mbox{He}}p\pi$ reaction (right).}
\end{figure}

The experimental method is based on a measurement of the excitation function
for the
chosen decay channels of the ${\mbox{He}}-\eta$ system and the search for a resonance-like structure below the ${\mbox{He}}-\eta$ threshold. The measurement was performed by applying a continuous variation of the beam momentum around the threshold\footnote{The continous change of the momentum of the COSY beam was already successfully applied to measurements 
of the $\eta$ and $\eta^{\prime}$ meson production in $pp$ and $pd$ 
reactions~\cite{jurek-he3,timo,jureketa,moskal-prl}}. 
The relative angle between the outgoing $nucleon-pion$ pair which  originates from the decay of the N*(1535) resonance created via  absorption of
the $\eta$ meson on a nucleon in the $\mbox{He}$ nucleus, is equal to 180$^{\circ}$ in the $N^*$ reference frame. It is smeared by about 30$^{\circ}$ in the center-of-mass frame (see Fig.~\ref{angle}) due to the Fermi motion of the nucleons inside the $\mbox{He}$ nucleus.
The center-of-mass kinetic energies of the nucleon and pion originate from the mass difference $m_{\eta}-m_{\pi}$ and are around 50 MeV and 350 MeV, respectively.

Fig.~\ref{angle} shows that the distribution of the relative $proton-pion$ angle
expected for the background due to the prompt $dd\to {^3{\mbox{He}}}p\pi$ reaction is much broader 
than the one expected from the decay of the bound state.
This will allow to control the background by comparing "signal-rich" and "signal-poor" regions. 

The experiment was carried out with the WASA-at-COSY detector, 
and with the internal deutron beam of COSY  
scattered on a deuteron pellet target. 
A detailed detector description can be found in~\cite{Wasa1,michal}. 
We used a slowly ramped beam  
momentum scanning the range of momenta corresponding to a variation of the 
excess energy for the $^4$He$\eta$ system from -51.4~MeV 
to 22~MeV. 
This range is about three times wider than the width of the structure 
observed by the TAPS collaboration at MAMI in the $\gamma {^3\mbox{He}} \rightarrow \pi^0 p X$ process 
($25.6 \pm 6.1$)~MeV. 
\begin{figure}[b]
\psfig{file=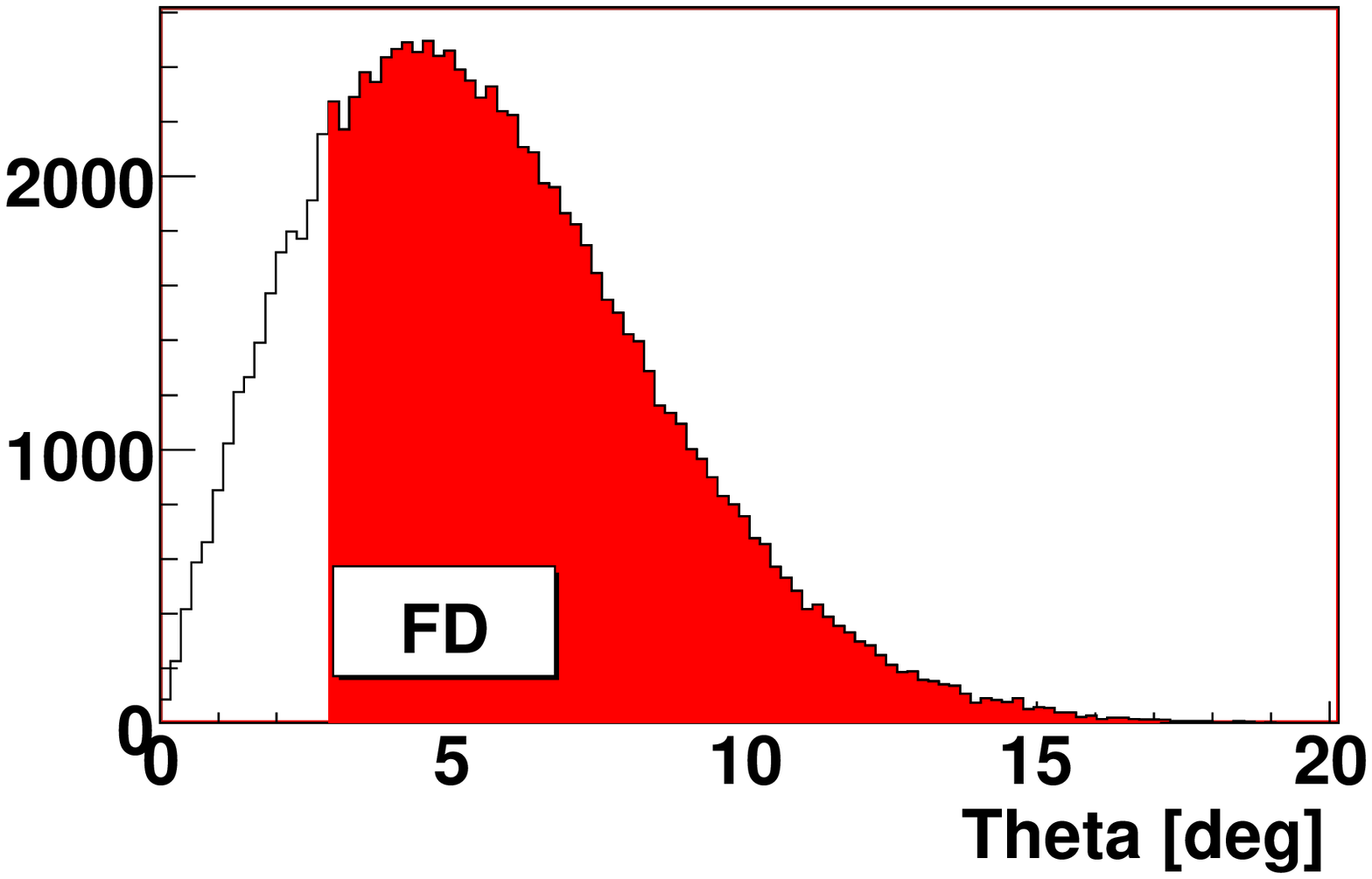,width=4.1cm}
\psfig{file=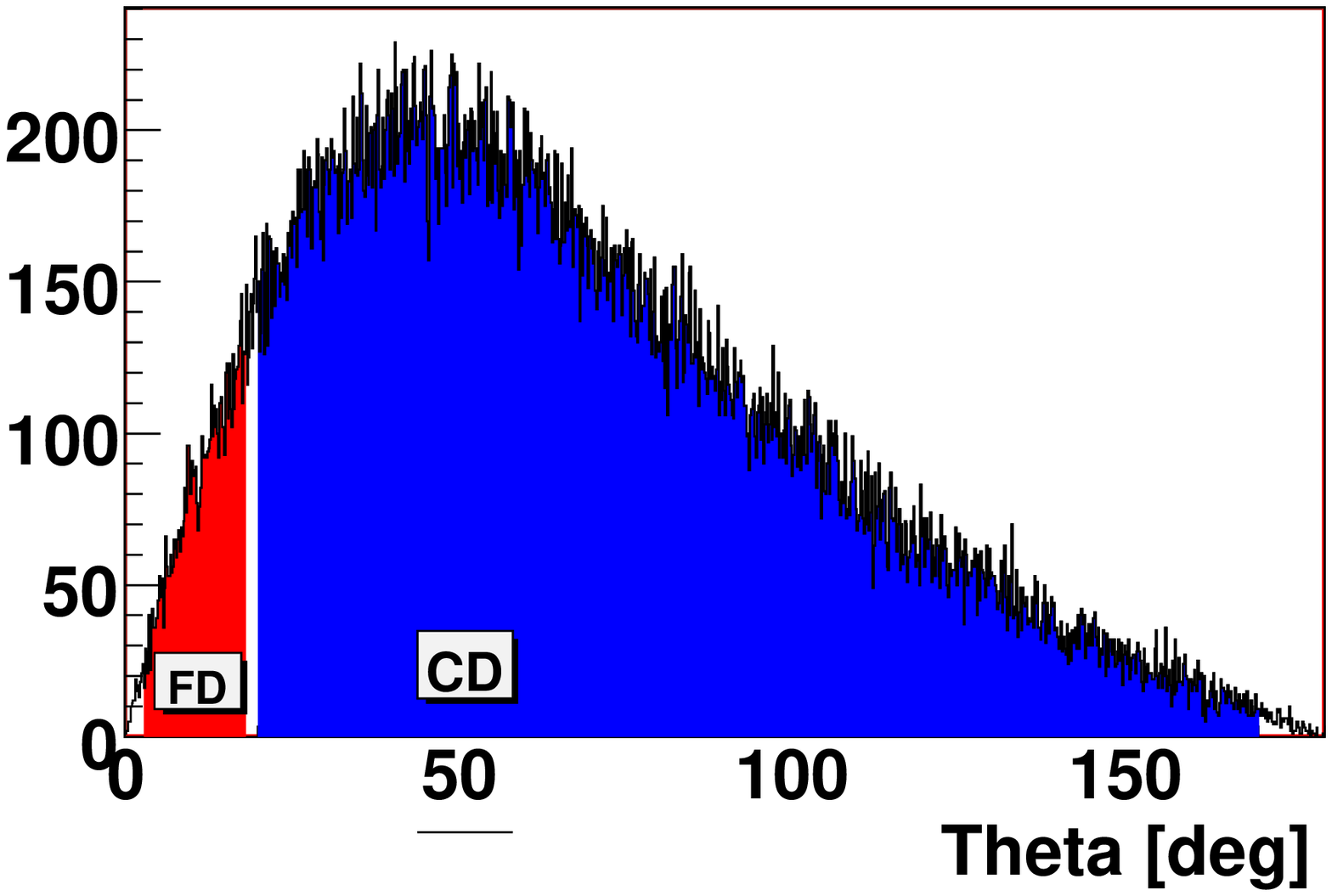,width=4.1cm}
\psfig{file=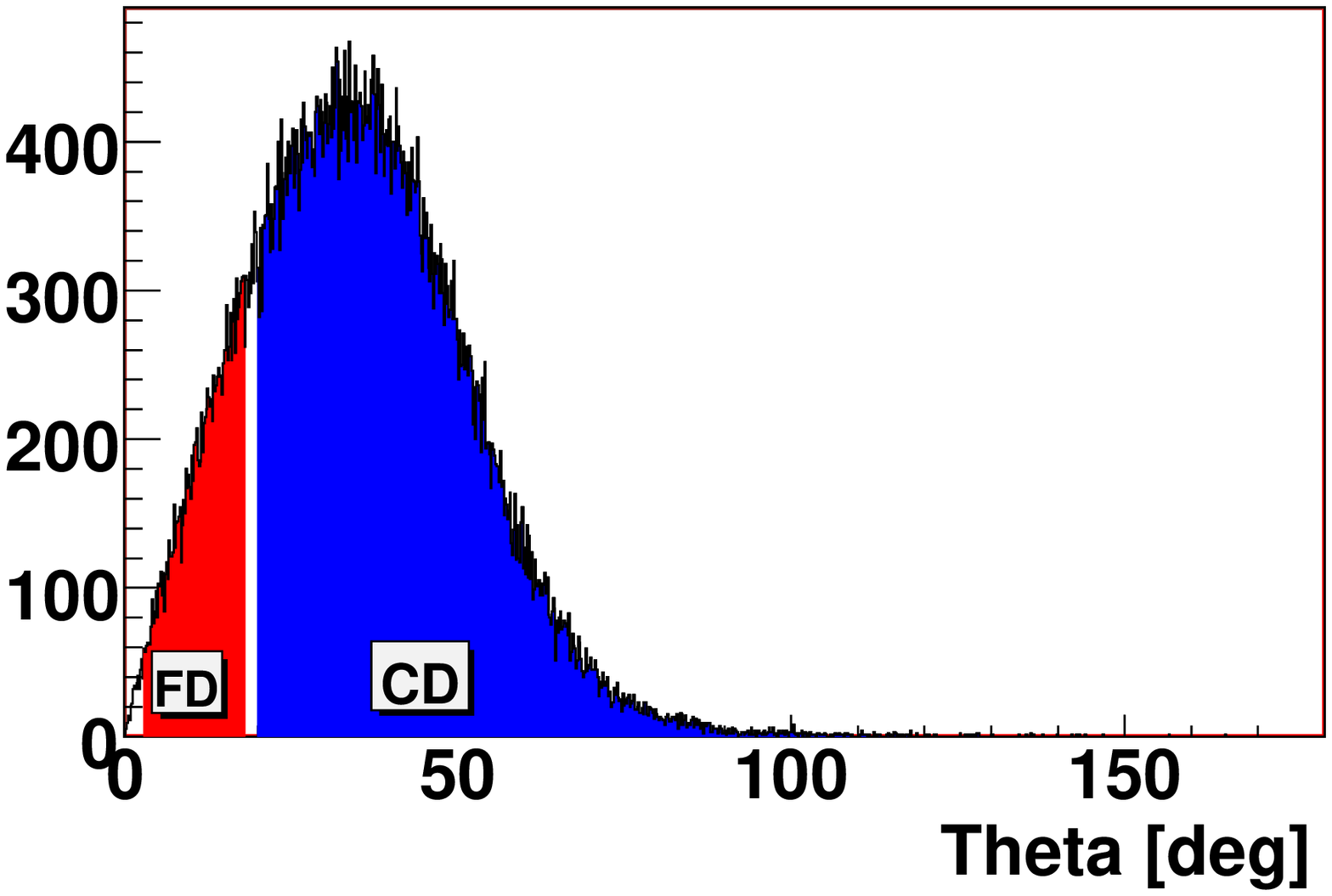,width=4.1cm}
\caption{\label{ptheta} Angular distributions  of $^3{\mbox{He}}$ (left), pions (middle) and protons (right) 
outgoing from 
the $dd \to (^4{\mbox{He}}\eta)_{bound} \to ^3{\mbox{He}} p \pi^-$  reaction. Labels  "FD" and "CD" indicate angular ranges 
covered by the Forward Detector and Central Detector, respectively. Please note that the presented range for $^3{\mbox{He}}$
ions is smaller than those for protons and pions.}
\end{figure}
The charged pions and protons are registered in the Forward Detector as well as in the Central Detector. ${^3\mbox{He}}$-ions are detected in the Forward Detector only.  
Fig.~\ref{ptheta} presents an example of the angular distributions of ejectiles expected for the  
$dd \to (^4{\mbox{He}}\eta)_{bound} \to ^3{\mbox{He}} p \pi^-$  reaction.  The simulations  
were conducted under the assumption that the decay of the bound state is caused 
by the decay of the $N^*$ into a proton $\pi^-$ pair leaving the $^3{\mbox{He}}$  
nucleus as a spectator with the Fermi momentum which it possessed inside the $^4{\mbox{He}}\eta$ nucleus. 
The Fig.~\ref{ptheta} shows that a significant fraction of the angular range of the outgoing particles 
is covered by the WASA-at-COSY detector setup. In particular, 
about 80\% of $^3{\mbox{He}}$ ions can be  
registered by means of the Forward Detector
(left panel of Fig.~\ref{ptheta})
and a coincident measurement of all ejectiles from $dd \to (^4{\mbox{He}}\eta)_{bound} \to ^3{\mbox{He}} p \pi^-$ reaction can be performed with an efficency of 70 \% which is flat in the whole range of the excess energy from -51.4~MeV to 22~MeV.

\section{First measurements}

In June 2008 a test measurement of the production of  
$p \pi^0 X$,  $p \pi^- X$   as well as of $T p \pi^0$ and  $^3{\mbox{He}} p\pi^-$ 
in the dd collisions with the WASA-at-COSY facility  was performed.  

\begin{figure}[h] 
\psfig{file=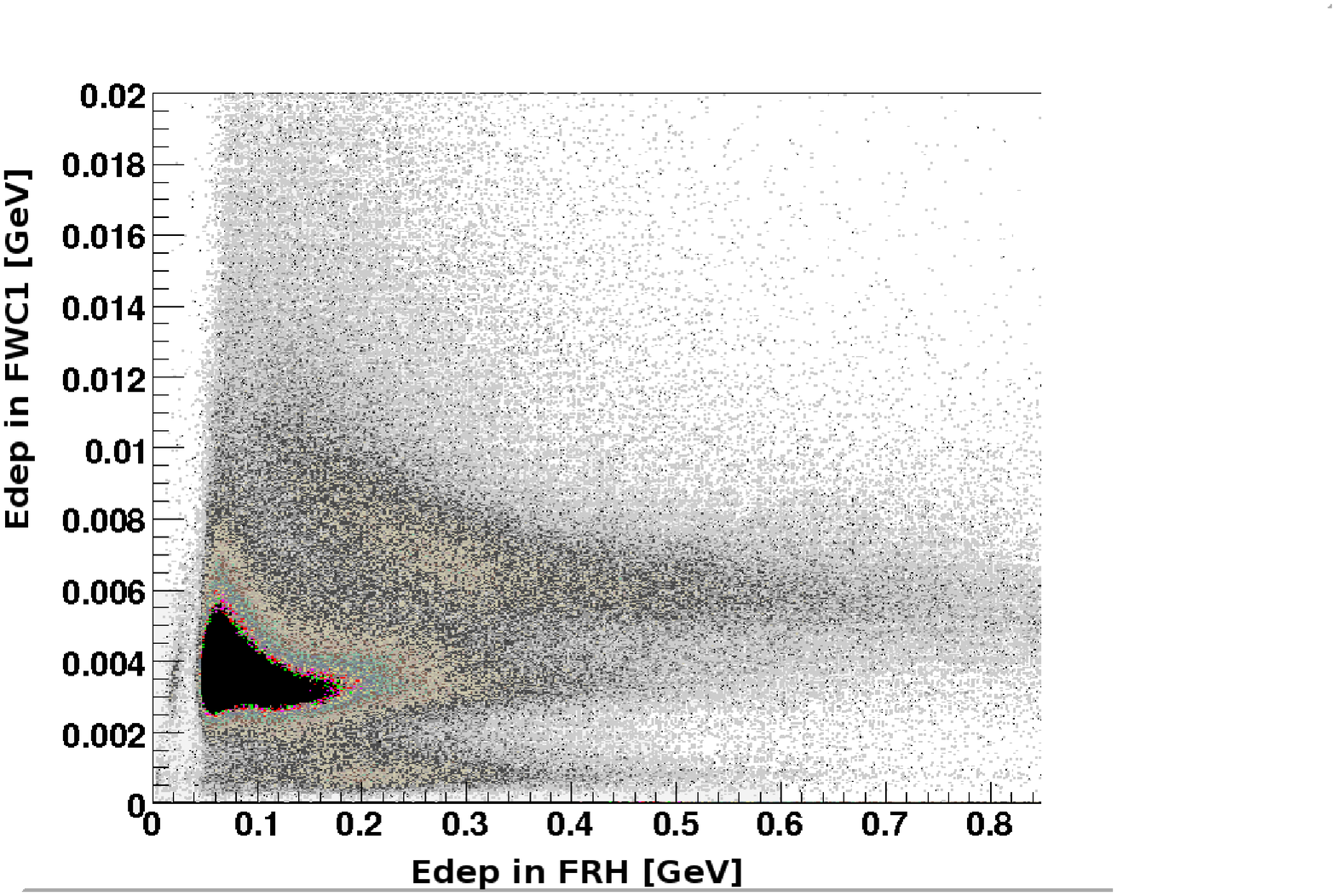, width=7.3cm}  
\psfig{file=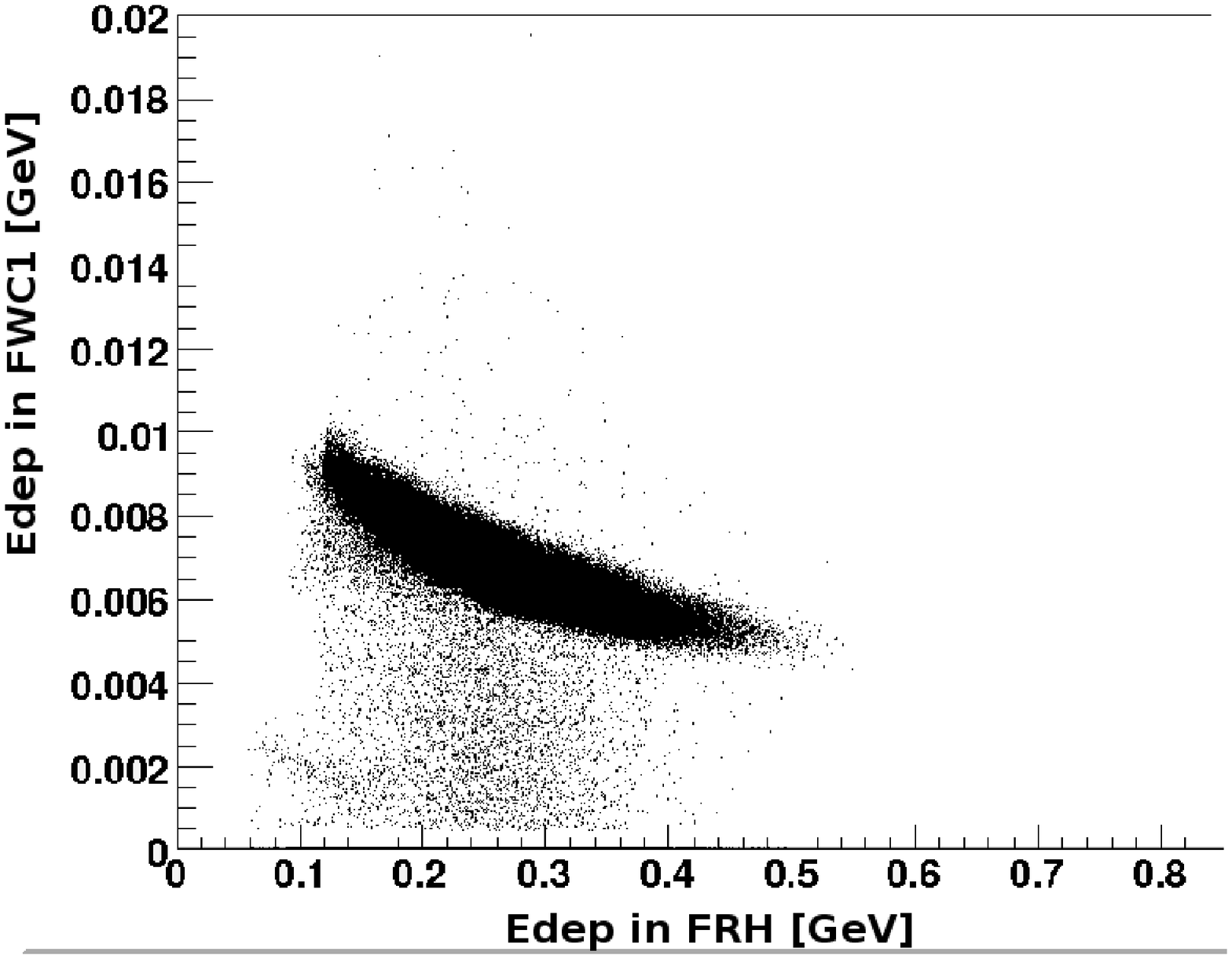,width=7.3cm} 
\caption{\label{de}  
 Energy loss in the Window Counter as a function of the energy loss in the Forward Range Hodoscope 
 as obtained in the experiment (left), and as simulated for the $^3{\mbox{He}}$ ions outgoing from the  
  $dd \to (^4{\mbox{He}}\eta)_{bound} \to ^3{\mbox{He}} p \pi^-$ reaction (right). 
} 
\end{figure}

The analysis of the energy signals from the Forward Detector revealed that it is possible 
to identify clearly the $^3{\mbox{He}}$ ions. In Fig.\ref{de} the experimental distribution of the energy loss in the Window Counter 
versus the energy deposited in the Range Hodoscope  
is compared to the expectation based on Monte-Carlo simulations. 

\begin{figure}[h] 
\centering
\psfig{file=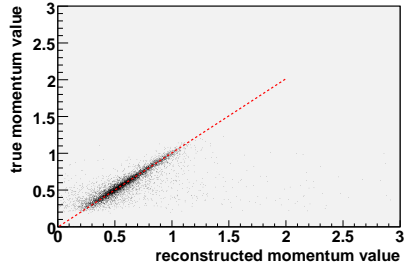,width=6cm}   
\caption{\label{momRecovsTrue} Monte Carlo:~Reconstructed momentum versus true momentum for two charged particles registered in the Central Detector under the condition that a
$^3\mbox{He}$
nucleus was detected 
in the Forward Detector} 
\end{figure}

\begin{figure}[h] 
\psfig{file=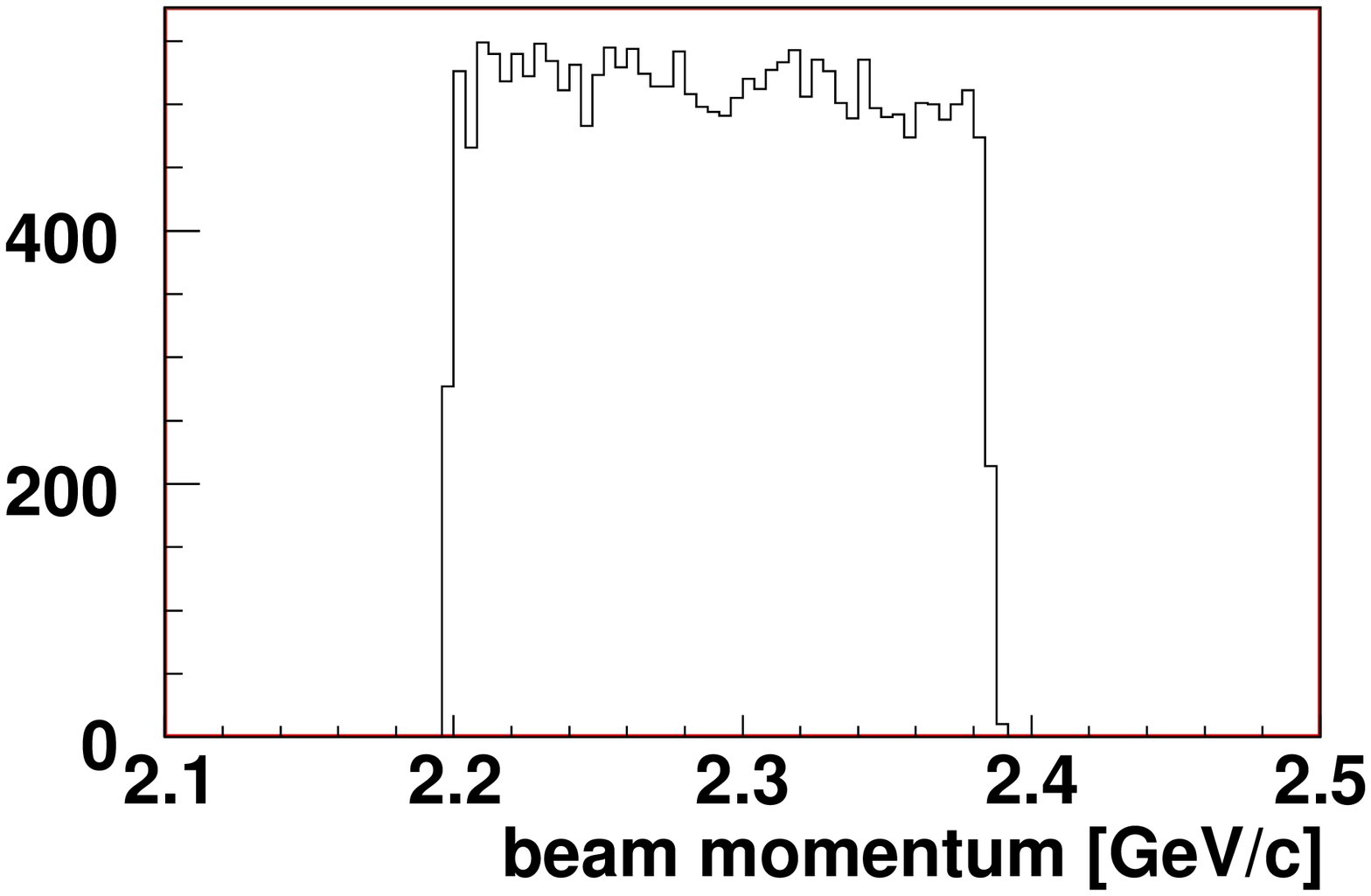,width=5.3cm} \hfill 
\psfig{file=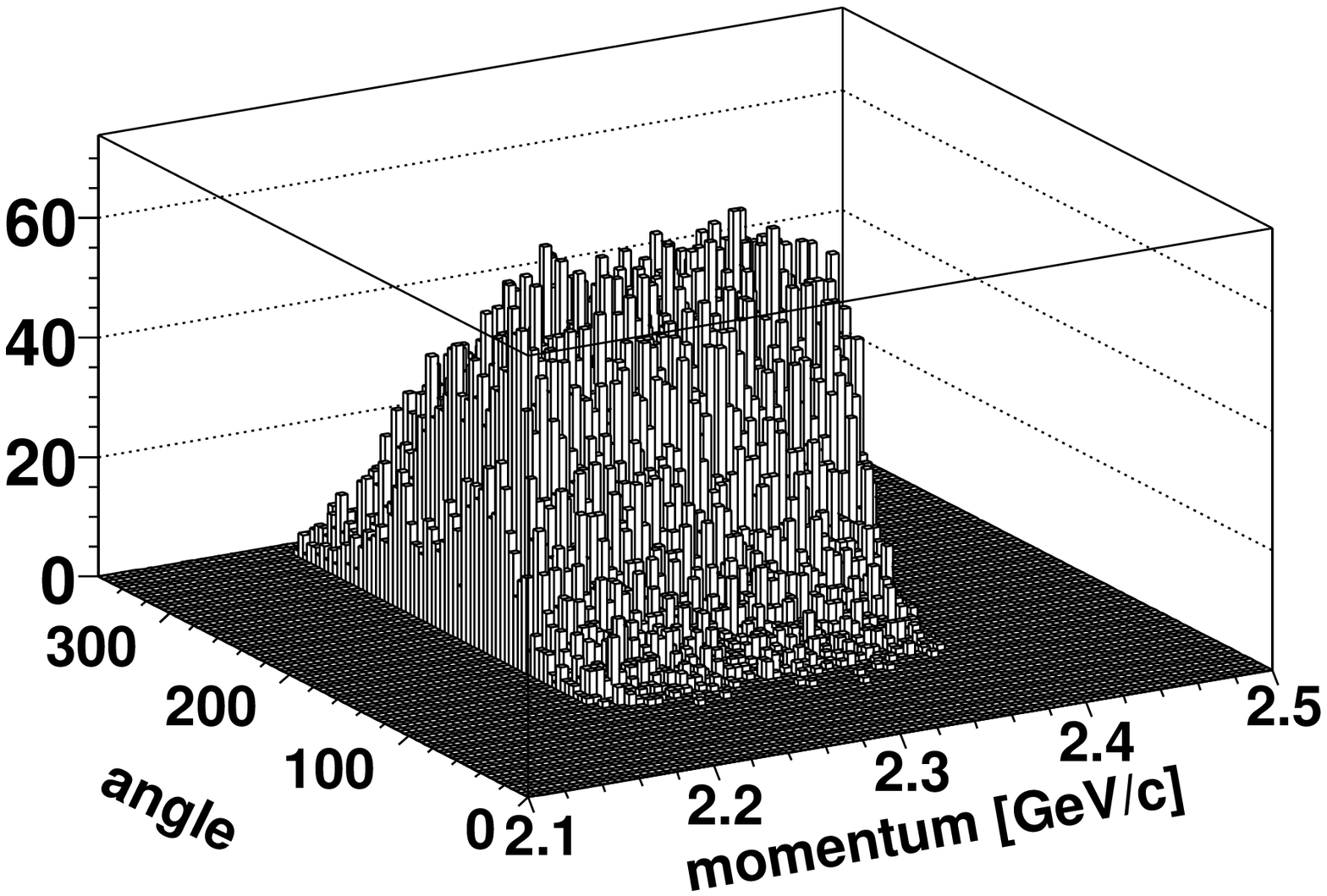,width=5.3cm}  
\caption{\label{excitations} Excitation functions for the $dd \to ^3\!{\mbox{He}}~(+two~charged-particles)$ as a function of the relative angle between the charged particles measured 
in the Central Detector (left).  Excitation function for the angular range between 160 and 200 degrees (right).} 
\end{figure}

Unfortunately, at the time of the experiment  
the cooling system of the superconducting solenoid was not working and we had to perform the measurement 
without magnetic field.  
That fact excluded the possibility of a direct momentum determination in the Central Detector as well as the use of the standard WASA-at-COSY identification method for the charged particles registered in the Central Detector. 
However, for the momentum determination of the charged particles in the Central Detector the information about directions of the particles in the Central
Detector, as well as, the
momentum of the $\mbox{He}$ nuclei can be used.
The absolute $^3\mbox{He}$ momentum  was determined on the basis of energy 
losses in the Forward Detector and its direction was reconstructed from signals in the set of drift chambers
built out of straw detectors. 
The directions of charged particles in
Central Detector
were extracted from signals in the 17 layers of the central drift chamber and from the Electromagnetic Calorimeter. 
Knowing the beam
momentum, the  $^3\mbox{He}$ momentum, and
the directions of the two other particles 
one can derive the momenta of pion and proton assuming energy and momentum conservation~(see Fig.~\ref{momRecovsTrue}). 
Energy loss in the Plastic Scintillator combined with the energy deposited in the Electromagnetic Calorimeter was used to identify protons and pions.
At present we have established the excitation function for the  
$dd \to ^3{\mbox{He}} (+two~charged~particles)$ reaction only. The result may be 
considered as a  
very conservative estimation of the upper limit for the prompt $dd \to ^3\!{\mbox{He}} p \pi^-$ reaction. 
The excitation function obtained is presented in Fig.~\ref{excitations}.  
Please note that  
the shape of the angular distribution (right panel) is consistent with the expectations shown in Fig.~\ref{angle}(right). 
In the left panel of Fig.~\ref{excitations}  the excitation function for the angular  
range between 160 and 200 degrees is presented. 

\section{Conclusions and Outlook}
We conduct a search for the ${\mbox{He}} - \eta$ bound state with the WASA-at-COSY facility. Because of the very high acceptance of the proton-pion pairs and good identification of helium nuclei the detector system  is very well suited for this kind of experiment.
In June 2008 we performed a first measurement of the excitation functions for the $dd \rightarrow {^3\mbox{He}}p\pi^-$ reaction.
The experiment will be continued in
the year 2010.
Two weeks of COSY beamtime  were already recommended by the COSY Program Advisory Committee. The use of the magnetic field will permit better proton-pion identification, as well as a better momentum determination in the Central Detector. 

In  two weeks of planned measurement, with the assumed luminosity of
\mbox{$4\cdot 10^{30}~$cm$^{-2}$ s$^{-1}$},
we expect to reach a sensitivity in the order of a few nb with a statistical significance of 1 $\sigma$. In the case of no observation of a signal this result will significantly lower the upper limit for the existence of a bound state.

\section{Acknowledgements}
The work was 
supported by the
European Community-Research Infrastructure Activity
under the FP6 program (Hadron Physics,RII3-CT-2004-506078), by
the German Research Foundation (DFG), by
the Polish Ministry of Science and Higher Education through grants
No. 3240/H03/2006/31  and 1202/DFG/2007/03,
and by the FFE grants from the Research Center J{\"u}lich.

\end{document}